\documentclass[11pt,reqno]{amsart}
\usepackage{curves} 
\usepackage{graphicx} 
\usepackage{amsmath, amsfonts, amssymb, amscd, soul, tikz}

\usepackage{amsmath, amsfonts, amssymb, amscd,
 enumerate, enumitem, color, xcolor, cancel,
graphpap,  mathrsfs
}
\theoremstyle{plain}
\newtheorem{theo}{Theorem}[section]

\theoremstyle{definition}
\newtheorem{rem}[theo]{Remark}

\theoremstyle{plain}

\theoremstyle{definition}

\renewcommand{\=}{:=}

\newcommand{\beq}{\begin{equation}}
\newcommand{\eeq}{\end{equation}}
\renewcommand{\a}{\alpha}

\renewcommand{\d}{\delta}

\newcommand{\g}{\gamma}
\newcommand{\h}{\eta}

\renewcommand{\l}{\lambda}
\renewcommand{\o}{\omega}

\renewcommand{\t}{\tau}

\newcommand{\bR}{\mathbb{R}}

\newcommand{\bW}{\mathbb{W}}

\newcommand{\cC}{\mathcal{C}}

\newcommand{\cG}{\mathcal{G}}
\newcommand{\cH}{\mathcal{H}}

\newcommand{\cK}{\mathcal{K}}

\newcommand{\cM}{\mathcal{M}}

\newcommand{\cS}{\mathcal{S}}

\newcommand{\cU}{\mathcal{U}}

\newcommand{\cW}{\mathcal{W}}

\newcommand{\p}{\partial}

\renewcommand{\square}{\kern1pt\vbox
{\hrule height 0.6pt\hbox{\vrule width 0.6pt\hskip 3pt
\vbox{\vskip 6pt}\hskip 3pt\vrule width 0.6pt}\hrule height0.6pt}\kern1pt}

\newcommand{\wt}{\widetilde}
\newcommand{\wh}{\widehat}

\newcommand{\be}{\begin{equation}}
\newcommand{\ee}{\end{equation}}

\def\<#1,#2>{\langle\,#1,\,#2\,\rangle}
\newcommand{\arr}{\begin{array}{rlll}}
\newcommand{\ea}{\end{array}}
\newcommand{\bea}{\begin{eqnarray}}
\newcommand{\eea}{\end{eqnarray}}
\newcommand{\bean}{\begin{eqnarray*}}
\newcommand{\eean}{\end{eqnarray*}}

\def\sideremark#1{\ifvmode\leavevmode\fi\vadjust{
\vbox to0pt{\hbox to 0pt{\hskip\hsize\hskip1em
\vbox{\hsize3cm\tiny\raggedright\pretolerance10000
\noindent #1\hfill}\hss}\vbox to8pt{\vfil}\vss}}}

\newcounter{ssig}
\newcounter{ttig}
\newcommand{\ve}{\varepsilon}

\title[Pontryagin Maximum Principle and Stokes Theorem]
{Pontryagin Maximum Principle \\ and  Stokes Theorem}

\author[F. Cardin and A.  Spiro]{Franco Cardin
\hskip 0.8cm Andrea Spiro}

\subjclass[2010]{49J15, 34H05}
\keywords{Maximum Pontryaging Principle; Mayer Problem; Stokes Theorem; Geometric Optimal Control}

\thanks{{\it Acknowledgments}. This research was partially supported by the Project MIUR ``Real and Complex Manifolds: Geometry, Topology and  Harmonic Analysis'' and by GNSAGA and GNFM of INdAM}

\begin{document}

\begin{abstract}
We present a new geometric unfolding  of a prototype problem of optimal control theory, the Mayer problem.  This approach is crucially based on the Stokes Theorem and yields to  a necessary and sufficient condition  that characterizes   the optimal solutions, from which the classical Pontryagin Maximum Principle is derived in a new insightful way. It also suggests  generalizations in diverse directions  of such famous principle. 
\end{abstract}
\maketitle
\section{Introduction}
The Pontryagin Maximum Principle (PMP) \cite{PBGM} is universally recognized as a point of arrival for the  modern calculus of variations, with great    achievements  in  both applied and pure  mathematics. All this  is clearly testified by   the vast literature on this subject -- see for instance  the  excellent  historical drawings that one can find in   \cite{Su1, Su2, Su3}. Thus,  it is quite unlikely that  further reconsiderations  of such celebrated  principle might determine  truly new    insights. Nonetheless this is precisely  what we try to do in this paper, being  confident  that our purely differential-geometric approach,  mainly built upon the  Stokes Theorem,    provides  a further understanding of the matter. \par
\smallskip
 The main ideas,   on which our presentation is based, are  simple and come from the differential geometric  approach to   variational principles.   First, one  has to  observe  that a  Mayer problem  for a controlled dynamical system is  equivalent  to determine the minimum  for the integral  of an appropriate  functional
 on  the curves that represent  controlled evolutions of the  system. Second,   one needs to  recall that the  Stokes Theorem  relates the  difference between the     integrals over   two  homotopic curves 
with  the value of   an appropriate double integral, computed  along the  surface that is generated by  the  homotopy that  joins the considered  two curves.  These   observations  yield   almost immediately   to  an interesting  necessary and sufficient condition on controlled evolutions to be  solutions to the  Mayer problem.  We call it   {\it Principle of Minimal Labour}. From such a principle,  the PMP and   various generalizations   can be  derived in a  simple way. \par
\smallskip
However,  in order to    carry out   the  outlined program,  some   auxiliary  steps need to be taken into account. In particular,  it is necessary to find: a) a convenient formalization  of  the notion of  ``optimization problem'', expressed in terms of a special class of curves in an appropriate manifold, particularly convenient for analyzing  the controlled evolutions of a   Mayer problems; b)  an encoding  of the notion of Pontryagin needle variations based on  such formalization. \par
\smallskip
These auxiliary steps  and the above described approach to the PMP are easily seen to be generalizable to optimization problems of different kind and  provide a new way  to deal with   them. In particular, they  indicate   
that the classical Mayer problems  belong to  a larger family of 
cost minimizing   problems for system under  constraints of variational type, a topic that we analyze in greater detail  in   \cite{CS1}.  They  also show  that  Pontryagin needle variations are related with (homotopic) variations of curves with  two parameters  and not with 
 just  one,  as it is customary considered   in the standard calculus of variations. They finally 
reveal the existence of an  intimate relation between the  PMP  
 and various  approaches  a la Poincar\'e-Cartan to  controlled dynamics.  
 To the best of our knowledge, it  is the first time in which all   such interesting and, at least to us, unexpected   issues  are   put in 
  an appropriate evidence. \par
\smallskip
Before concluding,  we would like to  recall  that dealing with homotopies and related 
objects is surely not   a new  idea in control theory nor in the  literature on  hyper-impulses. For instance,  it appears in the works of    Bressan and Rampazzo  on  ``graphic completions''   and ``control-completion'' (\cite{BR, CF2}). Further,   the infinitesimal version of PMP 
(which we obtain here as one of the possible consequences of  the  Principle of Minimal Labour) effectively consists of a system of differential equations for control systems that   are in a very strong relation with the equations of generalized Hamiltonian systems in  Tulczyjew's sense (see e.g. \cite{MT, Ca})  and  with the  equations of controlled Hamiltonian systems under  ideal constraints considered   by  Bressan \cite{Br1,Br2,Br3} and further studied in  \cite{Br,Ra1,Ra2, Ma1,Ma2, CF1, CF2}. We hope to  clarify  the exact terms of such  important relations  in a future work. 
 \par
 \smallskip
 The paper is structured as follows. In Sect.\,\ref{section2} and \ref{section3}, the above mentioned formalization of the  optimization problems   are given. In Sect.\,\ref{Section4} and \ref{section5},  we show how,  in Mayer problems, the Stokes Theorem allows to  compare  costs between pairs of   controlled evolutions and we derive   the Principle of Minimal Labour from this. In these two sections,   we provisionally consider  only  Mayer problems with smooth data and smooth controls.  Indeed, as it is   explained in  Sect.\,\ref{section2},   such a choice is made     for  letting   emerging in the most    neat way  the main ideas   of our  approach.  In Sect.\,\ref{Section6} and \ref{Section7}, we show that  the classical PMP  is a consequence of the  Principle  of Minimal Labour and we indicate how  our results  can be  improved  and become  applicable  to  a wider  class of 
   Mayer problems  with data of weaker regularity. In Sect.\,\ref{Section8}
  a few suggestions  for further developments are given.  
\par
 \medskip
\section{The basic ingredients of a control problem}
\label{section2}
\setcounter{equation}{0}
For the main purpose of   fixing  notation and terminology, we would like to begin our discussion  by   listing  
  the essential  ingredients, upon which the control problems we are interested in, as for instance the classical Mayer problem, are built. They are  the following five. 
\begin{itemize}[itemsep=5pt, leftmargin=10pt]
\item[$\bullet$] {\it A {\rm dynamical system} evolving in a manifold $\cM$ in dependence of  a real parameter $t$, varying in a fixed interval $[0,T]$}.\par
\vskip 0.2cm
For the examples  considered in this note, the manifold $\cM$  is  the standard phase space  $\cM=T^*\bR^{N}$.\par
\vskip 0.2cm
\item[$\bullet$] {\it A set   $\cK$, which we  call {\rm set of control parameters}}.\par
\vskip 0.2cm
The elements $U$ of such a set   might be of many  different types and might even have  unexpected characterizations.  An example  we  have in mind -- which is by far not the only possible one -- is given by the  pairs $U = (u(t), (a, b))$, formed by a 
continuous curve 
$u: [0,T]   \to K\subset \bR^M$   in some fixed space $K$ 
(\footnote{Here, we talk about continuous  curves only to avoid excessive technical details. In order to enlarge the class of control problems that can be analyzed with our approach,  to study,  one   might  surely consider   generalizations of   such classical  notion of  curves as,   for instance, (non-connected)  graphs of piecewise continuous functions.}) 
and  appropriate  initial data $(a = q(0), b = p(0))$  for curves $(q(t), p(t))$ in the phase space $\cM= T^*\bR^N$ (\footnote{What we are calling  here    ``set of control parameters'' should not be confused with the set $K \subset \bR^M$,  in which the curves $u(t)$, appearing just as first elements of the pairs $U \in \cK$,  take values. Unfortunately,  in the literature on control problems,  also the  set $K$ is often called ``set  of control parameters''.   We hope that such overlapping  terminologies would not be causes of confusion.}).\par
\item[$\bullet$] {\it A class   $\cG$ of  {\rm curves} of the manifold $[0,T] \times \cM$}.\par
\vskip 0.2cm

\vskip 0.2cm
In the basic examples of control problems we are going to consider,   the classical  Mayer problems,   the class  $\cG$ to be considered  is made of  the curves of with values in  $[0, T] \times T^* \bR^N$  of the form
\beq \label{0}
\g:[0,T] \to [0,T]   \times T^* \bR^N \quad \text{of the kind}\quad t \overset{\g}\longmapsto (t, q(t),p(t))
\eeq
The reason why one should   consider  such curves in the cartesian product $[0, T] \times T^* \bR^N$ will be shortly manifest, namely when  we will discuss costs, see  \eqref{alfa}  and \eqref{9}.\par
\item[$\bullet$] {\it A  well defined  {\rm correspondence} that  associates  with any   $U \in \cK$ a  unique well-defined curve   $\g^{(U)}$ of the class $\cG$. } \par
\vskip 0.2cm
For the Mayer problems, such  correspondence  comes from the  usual differential constraint 
\beq \label{-1}
\dot q=F(t, q, u(t)) , \quad q(0)=a\ ,
\eeq
 or, to be more precise, from 
 its  extended Hamiltonian  formulation, defined as follows.  For a given   constraint  of the form \eqref{-1}, consider   the function   (\footnote{Some authors prefer to  work  with the opposite function   $\wh \cH  = -p\cdot F(t, q, u)$ in place of $\cH =  p\cdot F(t, q, u)$ (see e.g. \cite{BC,Si}).   Our discussion  can be easily  developed  also using   such   $\wh \cH$, provided 
 that  a few  signs  in the definition of  the $1$-form  \eqref{18} are appropriately changed.
})
 \beq\label{1} \cH:[0,T] \times T^*\bR^N \times K  \to \bR\ ,\qquad 
  \cH(t, q, p, u) = p\cdot F(t, q, u) \ .
\eeq
The
 correspondence $\cK \to \cG$ that one has to use for  a Mayer problem   associates with  any pair 
$U = (u({\cdot}), (a, b))  \in \cK$ the unique curve $\g(t) = \g^{(U)}(t) =  (t, q(t), p(t)) \in \cG$,  which  is solution to  the differential problem 
\beq\label{2}
\begin{split}
 &\dot q = \frac{\p \cH}{\p p}\bigg|_{(t, q, p, u(t))} = F(t, q, u(t)) \ , \\
   & \dot p  = -  \frac{\p \cH}{\p q}\bigg|_{(t, q, p, u(t))} =-p\cdot \frac{\p F}{\p q}\bigg|_{(t, q, u(t))}\ ,\\
    & q(0) =a\ , \\
     &  p(0)  =b\ . 
      \end{split}
\eeq
As  is well known,  under appropriate standard assumptions of regularity (possibly relaxed a la Caratheodory or a la Filippov),  the Cauchy  problem \eqref{2}   has a unique solution and  such defined correspondence $\cK \to \cG$ satisfies  the  requirement of   being a well defined function (\footnote{As it is probably expected by  readers that are   familiar with the  basics of  classical control theory,  
the  differential  problem \eqref{2} will be shortly replaced by  an equivalent one,  in which  the  conditions 
$q(0) =a$ and  $p(0)  =b$ are replaced by  boundary conditions of the form
$q(0)=a$,  $p(T)=\bar b$.
This   replacement is   possible  due to   the  particularly simple structure  of the differential problem \eqref{2}, namely    by the fact that  the first equation $\eqref{2}_1$ is totally independent of  $p$.}). \\
\item[$\bullet$] {\it A  {\rm cost functional} 
\beq 
\label{I}
I: \{\ \g^{(U)}\in \cG \ ,\ U \in \cK\}  \longrightarrow \bR\ ,
\eeq
which assigns a well defined real number  (the {\rm cost}) to each of the curves $\g^{(U)}$ that are associated with the elements $U \in \cK$.}\par
\vskip 0.2cm
Assume  that  $\cK$,   $\cG$ and the correspondence $\cK \to \cG$ are as in the above described examples. Then,  given a   $1$-form  $\a$ of  
   $ [0,T] \times T^* \bR^N$  
 \beq
   \label{alfa}
\a = \a_0 dt + \a_i dq^i + \a^j dp_j\ ,
\eeq
 we may consider   the    cost functional $I_\a(\g^{(U)})$  defined by  
\beq
 \label{9}
 \begin{split}
I_\a(\g^{(U)}) &\= \int_{\g^{(U)}}\hskip -10pt \a = \\ & =\int_0^T \hskip-5pt
 \bigg(\a_0(\g^{(U)}(t)) + \a_i(\g^{(U)}(t)) \dot q^i(t) + \a^j(\g^{(U)}(t)) \dot p_j(t) \bigg) dt\ .
\end{split}
\eeq
We will shortly see that   for the classical Mayer problems, the cost functionals are precisely of this form.\par
\end{itemize}
\par 
This ends our list of   the five  ingredients we are considering for  the generic  ``control problems'', which we start discussing in the next section. \par
\smallskip
Before concluding this preliminary section, we would like  to add some very convenient   additional  convention.  Just for the  purpose  of  avoiding several  technical issues, in the next two sections
  {\it we   tacitly  assume  that $\cK$,  $\cG$, $\a$ and the correspondence $\cK \to \cG$ satisfy all possible additional   conditions,  which allow us the use of standard  calculus and classical differential geometric tools.}\par
\smallskip 
In other words,  we  assume  that all  data, needed to define the above five ingredients,  are 
differentiable  in the most  appropriate sense for making derivatives, integrals etc..  Moreover,  whenever it might be needed, we   assume that the set  $\cK$ is a path-wise topological space  and that all curves in $\cG$ are smoothly homotopic one to the other. \par
\bigskip
 A way to address the various  technical   issues,   which     arise under less  convenient (but more realistic)  assumptions,  is discussed in Sect.\,\ref{Section7}.
\par
\medskip
\section{What  a  control problem is }  
\label{section3} 
\setcounter{equation}{0}
Given a  dynamical system on $\cM$ and  the other    ingredients $\cK$, $\cG$, $\cK \mapsto \cG$ and $I$,   one can consider the following  general form   of  a control  problem.\par
\smallskip
\noindent {\bf  Problem.}  {\it  Determine which  elements  $U_o $  of a prescribed   subset   $\wt \cK  \subset \cK$ which  realize  the minimum for   the cost functional $I$ over the curves corresponding to the parameters in  $\wt \cK$, i.e. find the $U_o \in \wt \cK$ such that}
\begin{equation}\label{ast}
I(\g^{(U_o)}) \leq I(\g^{(U)}) \qquad \text{for all}\ \ \  U \in \wt \cK\ .
\end{equation}
The  elements $U_o$ that satisfy  (\ref{ast}) are called  {\it optimal solutions in the selected subset $\wt \cK$}.\par
\medskip
As mentioned in \S \ref{section2}, the main examples of control problems   we  want to consider  are the classical Mayer problems and are given by  dynamical system evolving in the phase space  $\cM = T^*\bR^N$ and such that: 
\begin{itemize}[itemsep=5pt, leftmargin=10pt]
\item[(a)] The sets $\cK$,  $\cG$ and the correspondence $\cK \to \cG$ are the set of pairs $U = (u(t), (a, b))$,  of curves $\g:[0,T]  \to [0,T] \times T^*\bR^N$ and the correspondence determined by the differential problem \eqref{2}, described in \S \ref{section2}; 
\item[(b)] The subset $\wt \cK \subset \cK$  is given by the collection of pairs $U = (u(t), (a, b))$,  in which  $a$ is  equal to a fixed values $a_o$. In this way,  the curves $\g^{(U)}$ corresponding to the elements  $U \in \wt \cK$ are just  the curves $\g^{(U)}(t) = (t, q(t), p(t))$,  in which $q(t)$ is solution to \eqref{-1} with initial value $a_o=q(0)$. This is precisely the class of motions that are considered  in  the classical Mayer problems. In \S \ref{section5}, the  arbitrariness on the second initial datum  $b = p(0)$ will be   determined by a convenient  condition on the final value $p(T)$.
\item[(c)]  the cost functional is  as in (\ref{alfa}), with  $1$-form $\a$ of the kind
\begin{equation}\label{18}
\a = p_j d q^j - \cH dt + \frac{\p C}{\p t}   dt +  \frac{\p C}{\p q^j} d q^j
  \end{equation}
 for some fixed smooth function  $C: [0,T] \times \bR^N \to \bR$, whose meaning will be clarified by  \eqref{24} below. We   assume that $C$, by construction, satisfies the condition
 \beq \label{3.3}
C(0, q) = 0\ .
\eeq
\end{itemize}\par
\bigskip
Working with such ingredients,  we see that 
at  each  point of the curve $\g^{(U)}(t)$ (which is solution to \eqref{2}) one has that 
 $$(p_j d q^j - \cH dt)(\dot \g^{(U)}_t) = p_j(t)( \dot q^j(t)  -   F^j(q(t), u(t), t)) = 0\ , $$
 so that,  on  each such   curve,  the cost functional  $I_\a$, defined in  \eqref{9}   is equal to 
  \begin{multline}\label{24}
 I_\a(\g^{(U)}) = \int_{\g^{(U)}} \a = \int_0^T  \left(\frac{\p C}{\p t}    +  \frac{\p C}{\p q^j} \dot q^j\right) dt  = \\
 = C(T, q(T)) - C(0, q(0)) = C(T, q(T))\ .
\end{multline}
This means that {\it     minimizing  the   cost functional \eqref{24}   amongst the  curves associated with  the control parameters in   $\wt \cK$,  is  equivalent  to minimize the value of the function   $C(T, q(T))$, amongst 
the values at the   final points   of the   curves $q(t)$, which  solve \eqref{-1} and have initial value $q(0) = a_o$}. \par
\smallskip
This is  usually described as the problem of 
{\it minimization of a   terminal cost under the differential constraint \eqref{-1}},  i.e. precisely what  is asked to do  in  a classical  Mayer problem.\par
\medskip
\begin{rem} \label{rem31} Being the terminal cost \eqref{24} completely independent of the function $p(t)$, 
{\it given an optimal   $U_o = (u_o(t), (a_o, b))$, also any  other pair $U'_o = (u_o(t), (a_o,b'))$ that  differs from $U_o$ only  by the  datum $b'$, is an optimal solution.}  In other words,  the  optimal solutions for the control problem determined by the class $\wt \cK$  are determined  up to arbitrary choices of the  datum $b = p(0)$. This very simple  observation will have  a crucial role in what follows.
\end{rem}
\par
\medskip
\section{Comparing costs  by means  of the Stokes Theorem}
\label{Section4}
\setcounter{equation}{0}
Let us take in action   a  Mayer problem  and the associated   problem described in previous section. We pick  two pairs   $U_o, U$ in  the  class  $ \wt \cK$  defined   in (b). \par
\smallskip
Faithful to our  convenient assumptions mentioned  at the end of  \S \ref{section2}, we assume that 
the subclass $\wt \cK$ is a  path-wise connected topological space, so that  we may consider a curve  $U(s)$, $s \in [0,1]$, in $\wt \cK$, with endpoints   $U(0) \= U_o$ and  $U(1) \= U$.  Since each  $U \in \wt \cK \subset \cK$ uniquely determines   a curve
$\g^{(U)}$  in  the class $\cG$,  the path  $U(s)$ in $\wt \cK$ uniquely determines a homotopy $\g^{(U(s))}$ of curves in $\cG$.  Moreover, being  the  considered curves  of the form  \eqref{0}, such homotopy is identifiable with a  continuous function  
$$\g = \g(t,s):[0,T] \times [0,1] \to [0,T] \times  T^* \bR^N\ ,$$
with the property that,   for each $s\in [0,1]$,  the map $ \g(\cdot, s) $ is the curve
\begin{equation}\label{6}
\g(\cdot, s) = \g^{(U(s))}(\cdot): [0,T] \to [0,T] \times T^* \bR^N\ ,
\end{equation} 
and, for $s = 0,1$ one has 
\begin{equation}\label{7}
\g(\cdot, 0)   = \g^{(U_o)}(\cdot) \ ,\qquad \g(\cdot, 1) = \g^{(U)}(\cdot)
 \end{equation}
\centerline{
\begin{tikzpicture}
\draw[fill]  (0.5, 0.2) circle [radius = 0.05];
\draw[fill]  (3.8, 0.7) circle [radius = 0.05];
\draw[fill]  (0.25, 2.3) circle [radius = 0.05];
\draw[fill]  (3.7, 2.85) circle [radius = 0.05];
\node at  (2, 2.85) {\tiny $\g^{(U)}$};
\node at  (2, 0.55) {\tiny $\g^{(U_o)}$};
\node [red] at  (2, 1.85) {\tiny $\g^{(U(s))}$};
\draw [ thin, blue] (0.5, 0.2) to [out=45, in=180] (2,0.85) to [out=0, in=207] (3.8, 0.7)  ; 
\draw [ ->, thick, blue] (2, 0.85) to  (2.02, 0.85)  ; 
\draw [ thin, blue] (0.25, 2.3) to [out=45, in=180] (1.5, 2.65) to [out=-5, in=207](3.7, 2.85)   ; 
\draw [ ->, thick, blue] (2, 2.61) to  (2.02, 2.61)  ; 
\draw [ thin, blue] (0.5, 0.2) to [out=65, in=290] (0.25, 2.3)  ; 
\draw [ ->, thick, blue] (0.52, 1.5) to  (0.515, 1.51)  ; 
\draw [ thin, blue] (3.8, 0.7) to   [out=155, in=270] (3.3, 1.7) to
 [out=90, in=310]  (3.7, 2.85)  ; 
\draw [ ->, thick, blue] (3.3, 1.7) to  (3.3, 1.71)  ; 
\draw [ thin, red] (0.56, 1.25) to [out=45, in=180] (1.5, 1.66) to [out=-5, in=207](3.3, 1.8)   ; 
\draw [ thin, red] (0.61, 0.75) to [out=45, in=180] (2, 1.26) to [out=-5, in=207](3.35, 1.15)   ; 
\draw [ thin, red] (0.44, 1.9) to [out=45, in=180] (2, 2.31) to [out=-5, in=207](3.63, 2.3)   ; 
\draw [ ->, thick, red] (2, 2.31) to  (2.1, 2.31)  ;
\draw [ ->, thick, red] (2, 1.26) to  (2.1, 1.26)  ;
\draw [ ->, thick, red] (2.2, 1.61) to  (2.3, 1.61)  ;
 \end{tikzpicture}
 }
 \centerline{\tiny \bf Fig. 1}
\vskip 1 cm 
 Given  $\g = \g^{(U(\cdot))}(\cdot)$ of this kind, it is useful to consider  (Fig. 2):
\begin{itemize}[itemsep=5pt, leftmargin=10pt]
\item[--] the  curves in $[0,T] \times T^* \bR^N$, described  by the endpoints of the curves $\g^{(U(s))}$ 
\begin{equation}\label{11}
\h^{(U_o, U|0)}(s) \= \g(0,s)\ \ \text{e}\ \ \h^{(U_o, U|T)}(s)\= \g(T,s)
 \end{equation}
\item[--] the $2$-dimensional submanifold  $\cS^{(U_o, U)}$  (\footnote{Without any additional assumption, the traces of the curves in the considered  homotopy might not determine a smooth $2$-dimensional submanifold. Nonetheless, as we explained at the end of \S \ref{section2}, for simplifying the discussion  we assume that the homotopy is sufficiently nice so that it does generate  a smooth surface.})
of   $[0,1] \times T^* \bR^N$, determined by the traces of the curves  $\g(\cdot,s)$, which is globally parameterized by the continuous map 
\begin{multline}\label{12}
\wh \cS^{(U_o, U)}: [0,T] \times [0,1] \longrightarrow  [0,T] \times T^* \bR^N\ ,\\
(t, s) \longmapsto \wh \cS^{(U_o, U)}(t, s) \= \g(t,s) = (t, q(t,s), p(t,s))\ .
 \end{multline}
 \end{itemize}
\vskip -0.2 cm
 \centerline{
\begin{tikzpicture}
\draw[fill]  (0.5, 0.2) circle [radius = 0.05];
\draw[fill]  (3.8, 0.7) circle [radius = 0.05];
\draw[fill]  (0.25, 2.3) circle [radius = 0.05];
\draw[fill]  (3.7, 2.85) circle [radius = 0.05];
\node [blue] at   (0, 1) {\tiny$\h^{(U_o,U|0)}$};
\node [blue] at   (4.1, 1.7) {\tiny$\h^{(U_o, U|T)}$};
\node [red] at  (2, 1.8) {$\cS^{(U_o, U)}$};
\draw [ thin, blue] (0.5, 0.2) to [out=45, in=180] (2,0.85) to [out=0, in=207] (3.8, 0.7)  ; 
\draw [ ->, thick, blue] (2, 0.85) to  (2.02, 0.85)  ; 
\draw [ thin, blue] (0.25, 2.3) to [out=45, in=180] (1.5, 2.65) to [out=-5, in=207](3.7, 2.85)   ; 
\draw [ ->, thick, blue] (2, 2.61) to  (2.02, 2.61)  ; 
\draw [ thin, blue] (0.5, 0.2) to [out=65, in=290] (0.25, 2.3)  ; 
\draw [ ->, thick, blue] (0.52, 1.5) to  (0.515, 1.51)  ; 
\draw [ thin, blue] (3.8, 0.7) to   [out=155, in=270] (3.3, 1.7) to
 [out=90, in=310]  (3.7, 2.85)  ; 
\draw [ ->, thick, blue] (3.3, 1.7) to  (3.3, 1.71)  ; 
\draw [ thin, red] (0.56, 1.25) to [out=45, in=180] (1.5, 1.66) to [out=-5, in=207](3.3, 1.8)   ; 
\draw [ thin, red] (0.61, 0.75) to [out=45, in=180] (2, 1.26) to [out=-5, in=207](3.35, 1.15)   ; 
\draw [ thin, red] (0.44, 1.9) to [out=45, in=180] (2, 2.31) to [out=-5, in=207](3.63, 2.3)   ; 
\draw [ ->, thick, red] (2, 2.31) to  (2.1, 2.31)  ;
\draw [ ->, thick, red] (2, 1.26) to  (2.1, 1.26)  ;
\draw [ ->, thick, red] (2.2, 1.61) to  (2.3, 1.61)  ;
 \end{tikzpicture}
 }
 \vspace{-0.3cm}
  \centerline{\tiny \bf Fig. 2}
\vskip 0.1 cm  
 By considering the standard  counterclockwise orientation of $\p \cS^{(U_o, U)}$ so that it can be considered as a positive cycle in $[0,T] \times T^* \bR^N$,  we  have   the following equality of chains
  \begin{equation}\label{clock}
 \g^{(U_o)} + \h^{(U_o, U|T)}  + 
 (-\g^{(U)}) +  (-\h^{(U_o, U|0)}) = {\partial \cS^{(U_o, U)} }\ .
\end{equation}
Thus, integrating the cost functional \eqref{alfa} along such a chain and {\it using  the  Stokes Theorem}, we have the following crucial identity
\begin{equation}\label{13}
 \hspace{-0.087cm}I_\a(\g^{(U_o)}) + \int_{\h^{(U_o, U|T)}} \a- I_\a(\g^{(U)})-\int_{\h^{(U_o, U|0)}} \a\, {=} \int_{\partial \cS^{(U_o, U)} } \a  \,{=}  \int_{\cS^{(U_o, U)} }d \a 
  \end{equation}
  \par
  \medskip
 Now,  in order to disclose the information encoded in \eqref{13}, it is convenient to introduce the following two notions.  Given the homotopy $s \overset \g \longmapsto  \g(\cdot, s)$ between the curves $\g^{(U_o)}(\cdot)$, $\g^{(U)}(\cdot)$ as  in (\ref{6}) e (\ref{7}),   define: 
  \begin{itemize}[itemsep=5pt, leftmargin=10pt]
  \item[$\bullet$]  The   {\it endpoints labour}    (\footnote{A much more natural name for this integral should be ``work''. We  chose the  name ``labour'' 
  for preventing confusions with such  classical mechanical object.})  as  the real number $\cC^{(U_o, U, \g)}$ given by 
\begin{equation}\label{14}
\cC^{(U_o, U, \g)} \= \int_{\h^{(U_o, U|T)}} \a - \int_{\h^{(U_o, U|0)}}\a
  \end{equation}
  \end{itemize}
$\bullet$ The   {\it $2$-dimensional labour}  as the   value $\cW^{(U_o, U, \g)}$  of the   double integral 
\begin{equation}\label{15}\cW^{(U_o, U, \g)} \= - \int_{\cS^{(U_o, U)} }d \a
  \end{equation}
By  \eqref{13}, the difference in   costs $\d I_\a = I_\a(\g^{(U)})- I_\a(\g^{(U_o)})$, between  the curves $\g^{(U)}$  and $\g^{(U_o)}$,   is  equal to 
$$\d I_\a =  \cC^{(U_o, U, \g)} + \cW^{(U_o, U, \g)} \ .$$
This immediately yields 
 to the following very simple, but  useful  fact: 
  {\it  the element $U_o \in \wt \cK$   is an optimal solution for the considered control problem {\rm if and only if }for  each  other  $U\in \wt \cK$ and for each homotopy  $\g = \g(t, s)$ in $\wt \cK$  between the curves  $\g^{(U_o)}$ and  $\g^{(U)}$, the sum of the endpoint labour $ \cC^{(U_o, U, \g)}$ and the $2$-dimensional labour $\cW^{(U_o, U, \g)}$ is always non-negative
\begin{equation}\label{16}  \cC^{(U_o, U, \g)} + \cW^{(U_o, U, \g)}   \geq 0 . \end{equation}}
\par
\medskip
\section{Labours in case of a classical Mayer problem: \\
the Principle of Minimal Labour}\label{section5}
\setcounter{equation}{0}
 We now   determine  the explicit expressions of the endpoint labours and the $2$-dimensional labours  for   the classical Mayer problem,   as it has been presented in \S \ref{section3},  i.e. with $\cK$, $\wt \cK$ and $\a$ defined in  (a), (b) and (c) of that section. \par
 \smallskip
Let us  first focus  on  the endpoint labour $\cC^{(U_o, U, \g)}$.  We   recall  that 
for any given homotopy $\g$ connecting two curves $\g^{(U)}$ and $\g^{(U_o)}$,  with  $U_o, U \in \wt \cK$ as  in (b) of \S \ref{section3},   the   curves   $\h^{(U_o, U|0)}$,  $ \h^{(U_o, U|T)}$ are   given by the endpoints of a one-parameter family of curves of   the form \eqref{0}.  In particular,   the projections of such curves onto  the $t$-axis are  either identically  equal to $0$ or identically  equal  to $T$. In both cases,  $dt\equiv 0$ along such curves of endpoints.  Hence,   for the 
$\a$ as in \eqref{18} and the set of pairs $(u, (a_o=q(0), b= p(0))) \in \wt \cK$ as  in (b), one has  
\begin{align}
\nonumber \cC^{(U_o, U,  \g)}& = \int_{\h^{(U_o, U|T)}}( p_j d q^j -\cH dt + \frac{\p C}{\p t}  dt +  \frac{\p C}{\p q^j} d q^j) - \\
\nonumber&  \hskip 2 cm - \int_{\h^{(U_o, U|0)}}(p_j d q^j - \cH dt + \frac{\p C}{\p t}dt +  \frac{\p C}{\p q^j} d q^j ) = \end{align}
\begin{align}
\nonumber&=   \displaystyle\int_{\h^{(U_o, U|T)}}( p_j d q^j  + \frac{\p C}{\p q^j} d q^j)\ \  -\hspace{- 0.4cm} \underbrace{\int_{\h^{(U_o, U|0)}}(p_j d q^j  +  \frac{\p C}{\p q^j} d q^j )\, ,}_{ \hspace{0.7cm}=\, 0, \ \text{since for any }s\in[0,1]:\  q(0,s)\equiv a} \\
 \label{espressioncina} & =   \displaystyle\int_{s\in [0,1], \g(T,s)}\Big(p_j(T,s) + \frac{\p C}{\p q^j}(T,q(T,s) )\Big)\frac{\p q^j}{\p s}(T,s) ds\ .
\end{align}
This gives an enlightening relation between    $ \cC^{(U_o, U,  \g)}$ and the functions  $ p_j(T,s) + \frac{\p C}{\p q^j}(T,q(T,s) )$ along  the  endpoints curve  $ \h^{(U_o, U|T)}$ at $t = T$. 
  \par
 \bigskip
 Let us now consider   the $2$-dimensional labour  $\cW^{(U_o, U,\g)}$. We start 
 by determining an  explicit expression of the differential $d\a$ along the points of curves $\g^{(U(s))}$, each of them  solution to the 
 differential problem \eqref{2}:
\begin{multline*} d \alpha = d(p_jdq^j-{\cH }dt)  =dp_j \wedge dq^j-({\cH}_{q^j} dq^j+{\cH }_{p_j} dp_j+{\cH }_{u^\ell} du^\ell)\wedge dt  \overset{dt\wedge dt =0}= \\
=
dp_j\otimes dq^j
-dq^j \otimes dp_j
-{\cH}_{q^j} dq^j \otimes dt +\\
+ 
{\cH}_{q^j} dt \otimes dq^j
-{\cH }_{p_j} dp_j\otimes dt+
{\cH }_{p_j} dt\otimes dp_j
 -{\cH }_{u^\ell} du^\ell\wedge dt = \\
\overset{\text{along solutions of \eqref{2}}} = - \underbrace{ ( {\cH}_{q^j} \dot q^j   +     {\cH}_{p_j} \dot p_j  )}_{=0}dt  \otimes dt- {\cH}_{u^\ell }d u^\ell \wedge dt =\\
 = -{\cH}_{u^\ell }\frac{\p u^\ell }{\p s}(t,s)ds\wedge dt = {\cH}_{u^\ell }\frac{\p u^\ell }{\p s}(t,s) dt \wedge ds
\end{multline*}
Using this expression, we see that  the $2$-dimensional labour  $ \cW^{(U_o, U,  \g)}$ (which, we recall,  is the integral of $d\a$ along the $2$-dimensional submanifold formed by the traces of solutions to \eqref{2}), reduces to
\begin{equation} \label{espressionab}
 \cW^{(U_o, U, \g)}{\begin{matrix}{}\\ {} =\\ {}^{(\ref{15})}\end{matrix}}- \int_{\cS^{(U_o, U)} }d \a= - \iint_{t\in [0,T], s\in [0,1]}{\cH}_{u^\ell }\frac{\p u^\ell }{\p s}(t,s)\, dt\wedge ds 
\end{equation}
\par
\bigskip
The identities    \eqref{espressioncina} and \eqref{espressionab} have some  interesting consequences.\par
\smallskip
First of all, the relation  \eqref{espressioncina}  suggests to consider a new   convenient   subclass of the (already restricted) set of controls $\wt \cK$.  In fact,  in our setting for the Mayer problem,  the collection $\wt \cK$  is given by the  
pairs $U = (u(t), (a = q(0), b = p(0)))$,  in which  $a$ is  fixed and equal $a_o$, but no restriction has been imposed on $b$.  We  may therefore consider the proper subset $\wt \cK' \subset \wt \cK$, given by the pairs $U =  (u(t), ( a_o = q(0),  b = p(0)))$  satisfying  the following property:  the unique solution $\g^{(U)}(t) = (t, q(t), p(t))$  to \eqref{2}, determined by the curve  $u(t)$ and the initial data  $(a_o = q(0), b = p(0))$,  is such that 
\beq \label{cond1bis}   p_j(T) = - \frac{\p C}{\p q^j}(T,q(T))\ .\eeq
Note that, in this way,  we restored   a very  familiar  condition in control theory. 
Due to the  simple form of  \eqref{2}, for each  initial datum  $ q(0) = a_o$ and each curve $u(t)$, there is a unique possible  $b$, such that the  solution with $p(0) = b$ satisfies \eqref{cond1bis}. It can be explicitly determined as follows:
\begin{itemize}
\item[--] solve  the first equation in \eqref{2}  with $q(0) = a_o$;  this is a problem not involving the unknown $p(t)$; 
\item[--] find the  $p(t)$ which solves  the second equation  with the boundary condition \eqref{cond1bis}. 
\end{itemize}
The initial value $b$, which  one is looking for,   is precisely    $b = p(0)$: \par
\smallskip
$$
\underset{\smallmatrix a_o=q(0)\text{ is fixed}\\ \text{and } b \text{\ is s.t.\ }\\(\ref{cond1bis})\ \text{is satisfied}
\endsmallmatrix}{\tilde \cK'}   \subset    \underset{\smallmatrix {\ \ }\\[5pt]a_o=q(0)\text{ is fixed}\endsmallmatrix}{\tilde \cK}   \subset        \underset{\smallmatrix {\ \ }\\[5pt] \text{\ no restrictions}\\ \text{on}\ 
 a\  \& \ b\endsmallmatrix}\cK \ .
$$
\smallskip
This smaller class $\wt \cK'$  is quite convenient, because due to \eqref{espressioncina} {\it for any homotopy $\g(t, s) = \g^{(U(s))}(t)$, determined by a curve $U(s) \in \wt \cK'$, the endpoint labour $C^{(U_o, U, \g)}$ is $0$}.  In this situation,  the differences between costs are  completely determined just by the  $2$-dimensional labour:
 \begin{equation}\label{16'}\d I_\a = I_\a(U)- I_\a(U_o)= \cW^{(U_o, U, \g)} \ . \end{equation}
\par
\medskip
Now, it is important to observe that,  if one replaces the original set $\wt \cK$ of control parameters with the proper subset $\wt \cK'$, from a purely formal point of view
  {\it   the new control problem is  different from the original one}: {\it the collection  of controlling data amongst  which one   looks  for the minimum  cost  is  now strictly smaller}. Nonetheless, it is also important to observe  that: 
\begin{itemize}[itemsep=5pt, leftmargin=12pt]
\item[$\bullet$]  If $U_o = (u_o(t), (a_o, b))$ is an optimal solution in $\wt \cK$  for the considered Mayer problem, due to Remark \ref{rem31},   also the  pair $U'_o = (u_o(t), (a_o , b'))$ with $b'$  so that \eqref{cond1bis} holds, is an optimal solution to the same   control problem.  Consequently, $U'_o$ is also an optimal solution to  the new control problem,  determined by  the  smaller set $\wt \cK'$ of control parameters. Let us call  such new optimal solution    {\it $p$-optimal} (\footnote{The name ``$p$-optimal'' has been chosen to remind that that it  differs from a generic optimal solution just for an appropriate change  of the initial  (and effectively, the  final) datum for  $p(t)$.}).  By these observations,  we may say that  {\it  up to a different choice of the  datum  $b = p(0)$,  each  optimal solution  corresponds to  a $p$-optimal solution and vice versa}. 
\item[$\bullet$]  By  \eqref{16'}, the   $p$-optimal solutions are characterized by the following easy
\end{itemize}
{\bf Principle of Minimal Labour.} {\it Necessary and sufficient  condition for an element  $U_o \in \wt \cK'$ to be a $p$-optimal solution  is that for any other  $U \in \wt \cK' $ and any homotopy  $\g= \g^{(U(\cdot))}(\cdot)$ in $\wt \cK'$, connecting the curves $\g^{(U_o)}$  and $\g^{(U)}$,   the associated $2$-dimensional labour  is non-negative, that is 
\beq \label{astast}  \cW^{(U_o, U, \g)} = - \iint_{t\in [0,T], s\in [0,1]}{\cH}_{u^\ell }\frac{\p u^\ell }{\p s}(t,s)\, dt\, ds  \geq 0\ .\eeq}\par
\par
\medskip
Combining these  two remarks, we can see  that  the above   Principle of Minimal Labour provides  a complete  characterization of  the optimal solutions to classical Mayer problems.
\par
\medskip
\section{The Pontryagin Maximum Principle as a consequence of 
the Principle of Minimal Labour}
\label{Section6}
\setcounter{equation}{0}
In this section, we   show how  the above Principle of Minimal Labour  can be used to derive  the Pontryagin Maximum Principle. Using the language of this notes, such classical principle   can be stated as  follows: \\[10pt]
{\bf Pontryagin Maximum Principle.} {\it Let  $U_o = (u_o(t), a_o, b) \in \wt \cK$ be  an optimal  solution to the considered Mayer problem. With no loss of generality,  we may assume it is $p$-optimal (see Remark \ref{rem31}).  Then  the associated curve 
$\g^{(U_o)}(t) = (t, q(t), p(t))$ is such that, 
for each   $\t \in [0, T]$ and  $\o \in K \subset  \bR^M$, 
\beq \label{PMP} \cH(\t, q(\t), p(\t), u_o(\t)) \geq \cH(\t, q(\t), p(\t), \o)\ . \eeq }
\par
\smallskip
By looking at   \eqref{astast},  one might be tempted to prove \eqref{PMP}    proceeding along the  following path. Given $\t \in [0,T]$ and $\o \in K \subset \bR^m$, consider a map $u^{(\t,  \o)}: [0, T]  \to K$, which is a  strongly  localized variation  of  $u_o(t)$ --  a  sort of $\d$-function --   equal to  $u_o(t)$  for  $t \neq \t$ and   equal to $\o$   at $t = \t$. After this, construct an homotopy $\g = \g^{(U(\cdot))}(\cdot)$, determined by a curve $U(s) \in \wt \cK'$ that  connects 
$U_o = (u_o(t), a_o, b)$ and $U = (u^{(\t, \o)}(t), a_o, b)$. Finally,  try to prove that, along such  homotopy $\g$, the integrand  in \eqref{astast} can be replaced by  the function $\frac{d}{ds}{\cH}$  and show  that the $2$-dimensional labour   takes the form
$$
-  \int_{ s\in [0,1]}\frac{d}{ds}{\cH} \ ds\Big|_{t=\tau} =   - {\cH}(\g^{(U_o)}(\tau), \omega) +  {\cH}(\g^{(U_o)}(\tau), u_o(\t))\ . $$
If one can  prove all this,  \eqref{PMP} would be just a simple  consequence of \eqref{astast}.  \par
\medskip
Such a  road-map is probably  correct, but  it cannot be easily pursued. One of the reasons is that  the above described  $\d$-function $u^{(\o, \t)}(t)$  cannot be considered as  a curve   in a traditional  sense.  Due to this,  in order to reach a rigorous proof, one should at first dramatically enlarge  the  class of  what, up to now, we are   calling    ``curves'', ``homotopies of curves''  and ``submanifolds generated by homotopies of curves''. Since our approach is  crucially rooted  on the Stokes  Theorem,  the whole project might really end up with   a rigorous proof only if also an appropriate generalization of  the Stokes Theorem is established. \par
\medskip
There is however another way to overcome all such technicalities   and  sophisticated preliminaries. It  is based on the use  of the so-called  {\it needle variations},    introduced by Pontryagin in his original proof  and which we   now  formulate in terms of  the language of this paper.\par
\medskip
As in the above statement of the PMP, let $U_o = (u_o(t), a_o, b) \in \wt \cK'$ be a  $p$-optimal  solution to the considered Mayer problem, and denote by 
$\g^{(U_o)}(t) = (t, q(t), p(t))$ the associated solution to \eqref{2}. Recall that, being $p$-optimal,  we also have that condition  \eqref{cond1bis} is satisfied.\par
\smallskip
Now, for each given $\t \in (0,T]$, $\o \in K \subset \bR^M$ and for each  sufficiently small $\ve > 0$, let us denote by 
$u_o^{(\t, \o, \ve)}: [0, T] \to K$  the piecewise continuous map 
\beq \label{uk}
u_o^{(\t, \o, \ve)}(t) \= \left\{\begin{array}{ll} u_o(t) & \text{if} \ t \in \big[0, \tau- {\ve} \big),\\[8pt]
 \omega & \text{if}\ t \in \big[\tau - {\ve},  \tau\big),\\[8pt]
 u_o(t) & \text{if} \ t \in \big[\tau, T\big]\ \\
 \end{array}\right.
 \eeq 
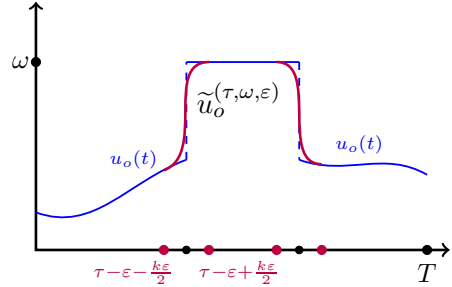
 and denote by $\wt u_o^{(\t, \o, \ve)}:  [0, T] \to K$ a smooth map, which appropriately approximates   $u_o^{(\t, \o, \ve)}$, i.e. it coincides with  it at all points with the only exception  of  two  ``very''  small neighborhoods of the discontinuities at  $t = \t - \ve $ and $\t = \t$ (\footnote{By ``very'' small neighborhoods, we mean open intervals, whose  width is less then or equal to $k \ve$ for some constant  $k < < 1$.}).  We call $u^{(\t, \o, \ve)}_o$  the {\it needle variation at $t = \t$ of ceiling value $\o$ and  width $\ve$}. Any associated continuous approximation  $\wt u_o^{(\t, \o, \ve)}$ will be  called {\it smoothed needle variation} (see Fig. 3 and Fig. 4). \par
 \medskip
 \centerline{
\begin{tikzpicture}
\draw[<->, line width = 1] (1,3.8) to (1,0.5) to (6.5,0.5);
\draw[<->, line width = 1] (7,3.8) to (7,0.5) to (12.5,0.5);
\draw[fill]  (3, 0.5) circle [radius = 0.06];
\node at  (3, 0.2) {\tiny $\t - \ve$};
\draw[fill]  (4.5, 0.5) circle [radius = 0.06];
\node at  (4.5, 0.2) {\tiny $\t$};
\draw[fill]  (6.2, 0.5) circle [radius = 0.06];
\node at  (6.2, 0.2) {\small $T$};
\draw[fill]  (9, 0.5) circle [radius = 0.05];
\draw[fill,purple]  (8.7, 0.5) circle [radius = 0.06];
\node[purple] at  (8.3, 0.2) {\tiny $\t {-} \ve{-} \frac{k\ve}{2}$};
\draw[fill,purple]  (9.3, 0.5) circle [radius = 0.06];
\node[purple] at  (9.7, 0.2) {\tiny $\t {-} \ve{+} \frac{k\ve}{2} $};
\draw[fill]  (10.5, 0.5) circle [radius = 0.05];
\draw[fill, purple]  (10.2, 0.5) circle [radius = 0.06];
\draw[fill,purple] (10.8, 0.5) circle [radius = 0.06];
\draw[fill]  (12.2, 0.5) circle [radius = 0.06];
\node at  (12.2, 0.2) {\small $T$};
\draw[fill]  (1, 3) circle [radius = 0.06];
\node at  (0.8, 3) {\small $\o$};
\draw[fill]  (7, 3) circle [radius = 0.06];
\node at  (6.8, 3) {\small $\o$};
\node[blue] at (2.3,1.7) {\tiny$u_o(t)$};
\node[blue] at (5.3,1.9) {\tiny$u_o(t)$};
\node[black] at (3.7, 2.5) {$u^{(\t, \o, \ve)}_o$};
\draw [line width = 0.7, blue] (1, 1) to [out=-20, in=200] (3,1.7)   ; 
\draw [line width = 0.7, blue, dashed](3,1.7)  to (3, 3)   ; 
\draw [line width = 0.7, blue] (3, 3) to  (4.5,3)   ; 
\draw [line width = 0.7, blue, dashed] (4.5,3)  to(4.5, 1.7)   ; 
\draw [line width = 0.7, blue] (4.5, 1.7) to [out=-20, in=150] (6.2,1.5)   ; 
\draw [line width = 0.7, blue] (7, 1) to [out=-20, in=200] (9,1.7)   ; 
\draw [line width = 0.7, blue, dashed](9,1.7)  to (9, 3)   ; 
\draw [line width = 0.7, blue] (9, 3) to  (10.5,3)   ; 
\draw [line width = 0.7, blue, dashed] (10.5,3)  to(10.5, 1.7)   ; 
\draw [line width = 0.7, blue] (10.5, 1.7) to [out=-20, in=150] (12.2,1.5)   ; 
\draw [line width = 1, purple](8.7,1.56)  to  [out=20, in=180] (9.3, 3) ; 
\draw [line width = 1, purple] (10.2,3)  to  [out=0, in=180] (10.8, 1.64)   ; 
\node[blue] at (8.3,1.7) {\tiny$u_o(t)$};
\node[blue] at (11.3,1.9) {\tiny$u_o(t)$};
\node[black] at (9.7, 2.5) {$\wt u^{(\t, \o, \ve)}_o$};
 \end{tikzpicture}
 }
   \centerline{\tiny\hskip 1 cm \bf Fig. 3 (Needle variation) \hskip 2 cm \bf Fig. 4 (Smoothed needle variation)}
 \smallskip
 \begin{rem}  A  rigorous derivation  of the classical PMP from our Principle of Minimal Labour should be built using  the smoothed approximations  $\wt u_o^{(\t, \o, \ve)}$ of the needle variations.
 However, with the   purpose of  being  as much as possible direct and clear, 
here  we  use    the (discontinuous)  needle variations $u_o^{(\t, \o, \ve)}$.  It is true that the  fact that the  $u_o^{(\t, \o, \ve)}$  are not $\cC^\infty$ can  make some of our  arguments sounding    not  completely right. Note however that everything  is immediately   fixed  by  just considering  smoothed needle variations in place of the discontinuous ones. 
\end{rem}
 \par
 \medskip
 Given a needle variation $u_\ve(t) \= u_o^{(\t, \o, \ve)}(t)$ of $u_o(t)$, we denote by 
 $$U_\ve = (u_\ve(t), (a_o, b_\ve))$$
  the unique pair  in $\wt \cK'$, with first element given by $u_\ve(t)$ and second element given by the pair  $(a_o, b_\ve)$ of initial values, chosen so that the corresponding curve $\g^{(U_\ve)}$  satisfies the condition \eqref{cond1bis} on the final value  $p(T)$. We also consider the homotopy $\g_\ve(t, s) = \g^{(U_\ve(s))}(t)$, where $U_\ve(s) = (u^{(s, \ve)}(t), (a_o, b^{(s, \ve)}))$ is the unique curve in the set $\wt \cK'$,  in which  
  \begin{itemize}
  \item[--] $u^{(s, \ve)}(t)$ is   defined by  (see Fig. 5)
\beq \label{uks}     u^{(s,\ve)}(t)\= 
(1 - s)  u_o(t) +  s u_\ve(t)\ .\eeq
\centerline{
\begin{tikzpicture}
\draw[<->, line width = 1] (1,3.8) to (1,0.5) to (6.5,0.5);
\draw[fill]  (3, 0.5) circle [radius = 0.06];
\node at  (3, 0.2) {\tiny $\t - \ve$};
\draw[fill]  (4.5, 0.5) circle [radius = 0.06];
\node at  (4.5, 0.2) {\tiny $\t$};
\draw[fill]  (6.2, 0.5) circle [radius = 0.06];
\node at  (6.2, 0.2) {\small $T$};
\draw[fill]  (1, 3) circle [radius = 0.06];
\node at  (0.8, 3) {\small $\o$};
\node[blue] at (2.3,1.7) {\tiny$u_o(t)$};
\node[blue] at (3.7,3.2) {\tiny$u_\ve(t)$};
\node[red] at (3.7, 2.5) {\tiny$u^{(s,\ve)}(t)$};
\draw [line width = 0.7, blue] (1, 1) to [out=-20, in=200] (3,1.7)  to   [out=20, in=160] (4.5, 1.7) to [out=-20, in=150] (6.2,1.5)   ;
\draw [line width = 0.7, red] (3,1.7)  to  [out=25, in=200] (3.5,2) to [out=10, in=180] (4, 2)  to [out=-10, in=140] (4.5,1.7);
\draw [line width = 0.7, red] (3,1.7)  to [out=30, in=200] (3.35,2.3) to  [out=5, in=180] (4.2, 2.3)  to  [out=-25, in=140] (4.5,1.7);
\draw [line width = 0.7, red] (3,1.7)   to [out=80, in=230] (3.2,2.7) to  [out=15, in=170] (4.3, 2.7)  to  [out=-50, in=100] (4.5,1.7);
\draw [line width = 0.7, blue, dashed](3,1.7)  to (3, 3)   ; 
\draw [line width = 0.7, blue] (3, 3) to  (4.5,3)   ; 
\draw [line width = 0.7, blue, dashed] (4.5,3)  to(4.5, 1.7)   ; 
 \end{tikzpicture}
 \hskip 3 cm 
 }
   \centerline{\tiny\bf Fig. 5 \hskip 2 cm }
   \item[--] the initial values $(a_o, b^{(s, \ve)})$ are chosen so that each  curve $\g_\ve(t, s) = \g^{(U_\ve(s))}(t)$, $s \in [0,1]$, satisfies the condition \eqref{cond1bis} on $p(T)$. 
\end{itemize}
Such homotopy $\g_\ve$, connecting the curves $\g^{(U_o)}$ and $\g^{(U_\ve)}$, uniquely determines the associated $2$-dimensional labour 
$\cW^{(U_o, U_\ve, \g_\ve)}$. \par
\smallskip
Now,  given  $\t \in (0,T]$ and  $\o \in K \subset \bR$ and setting $u_\ve(t) \= u_o^{(\t, \o, \ve)}(t)$,  
for a sufficiently small interval $(0, \bar  \ve)$,  we may  consider the  function 
$$W: (0, \bar \ve) \longrightarrow \bR\ ,\qquad W(\ve) \= \cW^{(U_o, U_\ve, \g_\ve)}\ .$$
We have that (Fig. 6): 
\begin{itemize}
\item[1)] The function $W$ is non-negative (by the Principle of Minimal Labour); 
\item[2)] The definition of $W(\ve)$ does not make any sense for  $\ve = 0$;  in fact, there is no possible smoothed variation $\wt u_o^{(\t, \o, \ve)}(t)$  when $\ve = 0$; nonetheless $W(\ve)$  can be extended  at $\ve = 0$ by setting $W(0) = \lim_{\ve \to 0}  W(\ve) = 0$; 
\item[3)] The function $W(\ve)$ is differentiable at all points $\ve > 0$, but, a priori, it might not be  differentiable at $\ve = 0$.
\end{itemize} 
\centerline{\hskip 3 cm 
\begin{tikzpicture}
\draw[ line width = 0.7, rounded corners, magenta] (1,3.8) to [out=-100, in=100] (1,0.5) to [out=-10, in=200] (6.5,0.5) to [out=80, in=-80] (6.5, 3.8) to [out=170, in=10] (1,3.8);
\draw[fill]  (5.5, 1.5) circle [radius = 0.05];
\node at  (5.5, 1.1) {$\g^{(U_o)}$};
\draw[line width = 0.5] (5.5, 1.5) to [out = 80, in = 260] (6.4, 2.4);
\draw[line width = 0.5] (5.5, 1.5) to [out = 100, in = 260] (5.5, 2.5);
\draw[line width = 0.5] (5.5, 1.5) to [out = 100, in = 280] (5.1, 2.55);
\draw[line width = 0.5] (5.5, 1.5) to [out = 110, in = 300] (4.5, 2.95);
\draw[->, line width = 0.5, dashed] (2, 3) to [out = -20, in = 190] (4.65, 2.5);
\draw[->, line width = 0.5, dashed] (2, 3) to [out = -20, in = 190] (5.13, 2.3);
\draw[->, line width = 0.5, dashed] (2, 3) to [out = -20, in = 190] (5.4, 2.15);
\draw[->, line width = 0.5, dashed] (2, 3) to [out = -20, in = 190] (6, 2.05);
\node at  (2.7, 3.2) {\tiny  \it homotopies $ \g_\ve = \{\g^{(U_\ve(s))}\}$};
\draw[ line width = 1, red]  (4.5, 3) to [out=-70, in=100] (6.7,2);
\draw [ ->, ultra thick, red] (5.7, 2.5) to  (5.71, 2.5)  ;
\node[red] at  (5.8, 2.9) {\tiny $\g^{(U_\ve)} = \g^{(U_\ve(1))}$};
\draw[fill, red]  (5.1, 2.56) circle [radius = 0.04];
\draw[ blue, ->, line width = 0.7]  (6, 2.55) to [out=10, in=130] (9,1);
\draw[->, blue, line width = 1]  (8,0.7) to (11,0.7);
\draw[fill, blue]  (8, 0.7) circle [radius = 0.05];
\node[blue] at  (8, 0.5) {\tiny $0$};
\node[blue] at  (11, 0.5) {\tiny $+\infty$};
\node[blue] at  (9.3, 2.3) {\small $W(\ve) \= \cW^{(U_o, U_\ve, \g_\ve)}$};
\node[magenta] at (3,1.4) {\small the space of curves $\cG$};
\draw[fill, red]  (6.68, 2.05) circle [radius = 0.05];
\node[magenta] at (7, 1.7) {\small $\displaystyle\lim_{\ve \to 0} \g^{(U_\ve)} $ };
\node[magenta] at (6.85,0.7) {\small $\p \cG $ };
 \end{tikzpicture}
 \hskip 3 cm 
 }
   \centerline{\tiny\bf Fig. 6}
     \par \medskip
However, by   (1), (2) and (3), it follows that {\it if  $\lim_{\ve \to 0} \frac{d W}{d\ve}\big|_\ve$ is proved to exists, then we have that $W$ is differentiable also at $0$ with non-negative derivative at $0$, i.e. }
$$ \frac{d W}{d\ve} \bigg|_{\ve = 0}  = \lim_{\ve \to 0} \frac{d W}{d\ve}\bigg|_\ve \geq 0\ .$$ 
The  limit $\lim_{\ve \to 0} \frac{d W}{d\ve}\big|_\ve$ does actually exist and it can be checked as follows.  First of all,   observe that for ant $\ve \in (0, \bar \ve)$, 
$$
\frac{\p }{\p s}u^{(s,\ve)}(t)= u_\ve(t)-u(t)=\left\{    
\begin{array}{ll}
 0 &  \text{if} \ t \in \big[0, \tau- {\ve} \big),\\[8pt]
 \omega -u(t) &\text{if}\ t \in \big[\tau - {\ve},  \tau\big),\\[8pt]
 0 &  \text{if} \ t \in \big[\tau, T\big].
\end{array}\right.
$$
Thus,  from \eqref{espressionab},
$$
\cW^{(U_o, U_\ve,\g_\ve)}=- \int_{s=0}^{s=1}\left(\int_{\tau-\ve}^\tau\left(  \omega^\ell-u^\ell(t)      \right)     
\frac{\p \cH}{\p u^\ell}\bigg|_{(t, q^{(s, \ve)} (t), p^{(s, \ve)}(t), u^{(s, \ve)}(t))} 
               dt  \right) ds
$$
and, for any fixed $\ve_o \in (0, \bar \ve)$ 
\begin{multline*}
\frac{d \cW^{(U_o, U_\ve,\g_\ve)}} {d\ve}\bigg|_{\ve_o}= \\
= - \int_{s=0}^{s=1}\left(   \omega^\ell-u^\ell(\t - \ve_o)      \right)     
\frac{\p \cH}{\p u^\ell}\bigg|_{\left(t, q^{(s, \ve_o)} (\t- \ve_o), p^{(s, \ve_o)}(\t- \ve_o), u^{(s, \ve_o)}(\t-\ve_o)
                 \right)}ds - \\
- \int_{s=0}^{s=1}\left(\int_{\tau-\ve_o}^\tau\left(   \omega^\ell-u^\ell(t)      \right) \right. {\cdot} \hskip 5 cm \\
{\cdot}   
\left. \frac{\p }{\p \ve}\left[    \frac{\p }{\p u^\ell}\cH\left(t, q^{(s, \ve)} (t), p^{(s, \ve)}(t), u^{(s, \ve)}(t)\right) 
 \right]_{\ve = \ve_o}
               dt  \right) ds\ .
\end{multline*}
It is now sufficient to observe that the second summand  in the right hand  side of this expression goes  to zero for $\ve_o \to 0$ (by the  Lebesgue Theorem on quasi-continuity) and  that 
$\lim_{\ve_o \to 0} \frac{d \cW^{(U_o, U_\ve,\g)}} {d\ve}\bigg|_{\ve_o}$ exists and is equal to
\beq\label{PMP1}
\begin{split}
\frac{d} {d\ve}\cW^{(U_o, U_{s},\g)}\big|_{\ve=0} &= - \int_{s=0}^{s=1}\left(   \omega^\ell-u^\ell(\t)      \right)     
\frac{\p \cH}{\p u^\ell}\bigg|_{\left(t, q(\t), p(\t), (1- s) u_o(\t) + s \o
                 \right)}\hskip - 1cm ds  = \\
                 &  =- \int_{s=0}^{s=1}    
\frac{\p}{\p s}   \cH ((t, q(\t), p(\t), (1- s) u_o(\t) + s \o)  ds = \\
                & = -  \cH(\t, q(\t), p(\t), \o) + \cH(\t, q(\t), p(\t), u_o(\t)) \ .
                 \end{split}
\eeq
Since we already observed that by the Principle of Minimal Labour, one necessarily has that  $\frac{d} {d\ve}\cW^{(U_o, U_\ve,\g_\ve)}\big|_{\ve=0} \geq 0$, from \eqref{PMP1}
the Pontryagin Maximum Principle follows. \square
\par 
\medskip
\section{Principles of Minimal Labour  
 in case of   non-smooth data} 
 \label{Section7}
\setcounter{equation}{0}
As it has been pointed out   in Sect.\,\ref{section2}, so far  we have just considered  classical Mayer 
problems that  are determined by  ingredients   $\cK$, $\cG$, $\cK \mapsto \cG$,  $I$ satisfying all    assumptions of smoothness and regularity that   render  the line of the arguments as much as possible straightforward.  It is now time to indicate   to which extent our discussion  can be generalised  to  Mayer problems   with lower regularity assumptions or  with additional   terminal  constraints.
\par
\smallskip
Consider  a  control  problem   in which  the ingredients $\cK$, $\cG$, $\cK \to \cG$,  $I$  are of  the following kind.\par
  \begin{itemize}
  \item The set  $\cK$ consists of pairs of the form $U = (u(t), (a,b))$, in which  $u(t)$ is  a $\cC^2$ function $u: [0,T] \to K$ taking  values into a  subset $K \subset \bR^M$, {\it which might be   disconnected or   with  empty interior}. 
  \item The class  $\cG$  consists  of $\cC^2$ curves  in $ [0,1] \times T^*\cM$  of the form \eqref{0}. 
  \item  For each  $U \in \cK$ the correspondence 
  $U \in \cK \mapsto \g^{(U)} \in \cG$
  maps $U = (u(t), (a,b)) $ into  the unique solution to the differential problem \eqref{-1} for some   $\cC^2$-function $F(t, q^i, u^\ell)$ which is   defined on  a set of the form $[0,1] \times \cM \times \cU_K$ {\it for some  convex neighbourhood  $\cU_K \subset \bR^M$  of $K$}. 
  \item  The cost functional  $I$ has the form (\ref{alfa})  with  $\a$ as in  \eqref{18}  with  $ C(t,q)$ of class $\cC^2$ on $[0,1] \times \cM$ and satisfying  \eqref{3.3}. \par
  \end{itemize}
  Let us also denote by  $\wt \cK$  the subclass of $\cK$  defined  in (b) of Sect.\,\ref{section3}, so that   the  control problem,   which is determined by  $\wt \cK$,  is precisely a classical Mayer problem on the curves $\g^{(U)}(t) = (t, q(t), p(t))$ with the initial value $q(0)  = a_o$.\par
  \smallskip
The above regularity assumptions on the   $u(t)$, $F(t, q, u)$ and  $C(t, q)$ lead to the following  fact.  Consider two curves $u_o(t)$ and $u(t)$ with values in $K \subset \bR^m$ and  pick a $\cC^2$-homotopy of curves    $u(t,s)$ such that: \\
\phantom{aa}  a) it  varies  between  $u_o(\cdot) = u({\cdot}, 0)$ and $u(\cdot) = u({\cdot}, 1)$;\\
\phantom{aa} b) it   takes  values in $\cU_K$ (but not necessarily  just in $K$). \\
Then, let  $U(s)$, $s \in [0,1]$, be  a curve of pairs of the form   $U(s)= (u(\cdot, s), (a_o, b(s)))$, with $b(s)$  of class $\cC^2$, and   set  $U_o \= U(0)$ and $U \= U(1)$.  Notice that, by construction, the endpoints $U_o$,  $U$  of the curve $U(s)$ are both  in  $\wt \cK$, but for $s \neq 0, 1$ the pairs     $U(s)$  are   not necessarily in $\wt \cK$. \par
\smallskip
These   assumptions imply  that: 
 \begin{itemize}[leftmargin = 30pt]
 \item[(1)] The surface $\cS^{(U_o, U)} \subset [0,T] \times T^* \cM$,  spanned  by the (traces of the) solutions $\g^{(U(s))}$  to \eqref{-1} with  data $U(s) = (u(t, s), (a_o, b(s)))$,   is  actually the image of a $\cC^2$-embedding $\wh \cS^{(U_o, U)}: [0,T] \times [0,1] \to  [0,T] \times T^*\cM$; 
 \item[(2)] All  coefficients of the $1$-form $\a$ are   of class $\cC^1$; 
 \item[(3)] The identity \eqref{13} holds also   in this situation. 
 \end{itemize}
 The claim (3)   follows from the fact that  the  Stokes Theorem  is still  valid  under the   regularity properties  (1) and (2) (see e.g. \cite[Ch.\,20, \S 6]{La}). 
 \par
 \smallskip
Due to this, the same   arguments considered in  Sect.\,\ref{Section4} immediately yield to the following: 
    {\it  an element $U_o \in \wt \cK$   is an optimal solution for the considered Mayer problem  if and only if for  any  other  $U\in \wt \cK$ and for all homotopies  $\g = \g(t, s)$ between the curves  $\g^{(U_o)}$ and  $\g^{(U)}$, generated by  homotopies $u(t, s)$ with values  in the convex open set $\cU_K$ ({\rm  but not necessarily just in  $K$})   with the above described regularity,  the sum between  the endpoint labour $ \cC^{(U_o, U, \g)}$ and  the $2$-dimensional labour $\cW^{(U_o, U, \g)}$ satisfies }
\begin{equation}\label{16bis}  \cC^{(U_o, U, \g)} + \cW^{(U_o, U, \g)}   \geq 0 . \end{equation}
If we further restrict  $\wt \cK$ to the proper subset $\wt \cK'$ that  is  determined by   the terminal condition \eqref{cond1bis} 
and  if we constrain  the  homotopies  to be made of  curves, determined by  pairs  $U(s) = (u(t,s), (a_o, b(s)))$ that are all  satisfying  \eqref{cond1bis}, then   \eqref{16bis} simplifies into $\cW^{(U_o, U, \g)}   \geq 0$. This gives  
{\it a  version of the Principle of Minimal Labour  holding under the above described  weaker regularity assumptions and in case of 
a set $K \subset \bR^M$, which may be disconnected  or with empty interior}.\par
 \smallskip
 The fact that $K$ is not required  to be connected or with non-empty interior is not truly  surprising.  In fact,  the  hypotheses  that  allows to derive   the Principle of Minimal Labour  are   basically just  the openness and the convexity (or, more generally, the path connectedness) of  the open set $[0,1] \times \cM \times \cU_K$,  where  the function $F(t, q^i, u^\ell)$    is well defined   and  of class  $\cC^2$.   \par
\medskip
 Starting now from this new    version of the Principle of Minimal Labour, the  previous proof of the Pontryagin Maximum Principle is still working. In this way we get the  PMP also  in case of a (possibly disconnected) set $K$ and with the above mild regularity assumptions on  $F(t, q, u)$, $C(t,q)$ and   $u(t)$.\par
 \smallskip 
A further weakening of the regularity assumptions can be very likely reached as  follows.   Instead of requiring that each  curve $u: [0, T] \to K$ is $\cC^2$, we may just impose that it is  piecewise $\cC^2$-regular, i.e. that it is a possibly discontinuous function  for which there is  a finite number of points $t_j \in [0, T]$  with the property  that  
  each restriction $u|_{(t_i, t_{i+1})}$ is $\cC^2$ and $\cC^2$-extendible  to the closed interval $[t_i t_{i+1}]$.  In this case,  even if the  surfaces $\cS^{(U_o, U)}$ that  are spanned by the usual homotopies of curves $\g^{(U(s))}$ cannot be expected to be  $\cC^2$,  it is  reasonable to predict  that they are  nonetheless  finite unions of $\cC^2$-surfaces, provided of course that the  function $F(t, q^i, u^\ell)$, which appears in \eqref{-1}, 
   is  of class $\cC^2$.  Using such a  decomposition in $\cC^2$-pieces, one can  still apply the  Stokes Theorem to   each $\cC^2$  piece of the surface and obtain  the inequality \eqref{16bis} also in these cases. Summing up,   we expect   that   the Principle of Minimal Labour can be obtained  and yield to  a   proof of  the  PMP  also under such weaker assumptions.  
   \par
 \medskip
  Variants of the Principle of Minimal Labour  with  weaker regularity  are  out of reach  if we limit ourselves to the above quoted  version of the  Stokes Theorem. 
 \par
 \smallskip
On the other hand,  it is very well known  that the PMP  has already been  proved  in  more   general settings,  as for instance those in which   the  $u(t)$   are  merely  bounded and  measurable  and  $F(t, q^i, u^\ell)$  is
just continuous (see e.g.\ \cite[Ch.\,6]{BP} for precise statements).  We believe that  these more  general versions can be 
deduced from the above ``piecewise smooth" PMP  using  approximations.  A full clarification of this expectation would be   useful and we hope to address it in a future work. It would pave the  way to a systematic two-step use of the Principle of Minimal Labour:  first, begin with  piecewise smooth data and infer from  the Principle of Minimal Labour  not only the PMP but  also  statements  of other     necessary  conditions   on the  optimal solutions (see,  for instance,  those   hinted  in  Sect.\,\ref{suggestions});    second, find a way to prove the new statements also in  case  of   weaker regularity  using  approximations.
\par
\medskip
\section{Further   consequences  of the Principle of Minimal Labour}
\setcounter{equation}{0}
\label{Section8}
\subsection{The labour functional}
\hfill\par
The Principle of Minimal Labour admits the following equivalent presentation. Let $U_o = (u(t), (a_o, b))$ be a pair in the restricted class $\wt \cK'$. As usual, for any other $U \in \wt \cK'$ let us consider an  homotopy of curves $\g(t, s) = \g^{(U(s))}(t)$, determined by a curve $ U(s) \in \wt \cK$ connecting $U_o$ with $U$. Note that, for each  choice of a point  $U_\l $, $\l \in [0,1]$,  of the curve $U(\cdot) \in \wt \cK'$, it is possible to rescale the parameter and obtain in this was  a new homotopy,
having the curves corresponding to $U_o$ and $U_\l$   as  endpoints:  
\beq \label{reducedspeed} \g_\l: [0, T] \times [0,1] \to [0, T]  \times T^* \bR^N\ ,\qquad \g_\l(t,s) \= \g(t, \l s)\ .\eeq
In this way, each  pair $(U, \g)$,  formed by a fixed $U \in \wt \cK'$ and a homotopy $\g(t, s) = \g^{(U(s))}(t)$,  connecting $\g^{(U_o)}$ and $\g^{(U)}$,  automatically determines   an entire new one-parameter family of 
pairs $(U_\l, \g_\l)$, formed by    $U_\l = U(\l)$ and  the homotopy, which is  defined in   \eqref{reducedspeed} and  connects  the curve $\g^{(U_o)}$ with  the curve $\g^{(U_\l)}$. 
It goes without saying that, conversely, a one-parameter family of 
pairs $(U_\l, \g_\l)$, $\l \in [0,1]$,  as above,  where the homotopies have the form 
\beq \label{donorione}  \g_\l(t,s) = \g_1(t, \l s)\qquad \text{for each}  \ \l \in [0,1]\ , \eeq 
is uniquely  determined by  final pair $(U, \g) \= (U_{\l = 1}, \g_{\l = 1})$, corresponding to the value $\l = 1$.\par 
This simple observation suggests that,  for each pair $(U, \g)$ as above, one can consider the function
\begin{multline*}\bW: [0, 1] \to \bR\ ,\\ \bW(\l) \= \cW^{(U_o, U_\l, \g_\l)}  =\hskip 8 cm  \\
=  - \iint_{t\in [0,T], s\in [0,1]}{\cH}_{u^\ell }\big|_{(\g^{(\l s)}(t), u^(t, \l s))}\frac{\p u^{\ell}(t, \l s)}{\p s}\bigg|_{(t,s)} dt\, ds = \\
= - \l \iint_{t\in [0,T], s\in [0,1]}{\cH}_{u^\ell }\big|_{(\g^{(\l s)}(t), u(t, \l s))} \frac{\p u^{\ell}(t, s')}{\p s'} \bigg|_{(t, s' =  \l s)}dt\, ds  \ .\end{multline*}
\centerline{\hskip 3 cm 
\begin{tikzpicture}
\draw[ line width = 0.7, rounded corners, magenta] (1,3.8) to [out=-100, in=100] (1,0.5) to [out=-10, in=200] (6.5,0.5) to [out=80, in=-80] (6.5, 3.8) to [out=170, in=10] (1,3.8);
\draw[fill]  (5.5, 1.5) circle [radius = 0.05];
\node at  (5.5, 1.1) {$\g^{(U_o)}$};
\draw[line width = 0.5] (5.55, 1.5) to [out = 108, in = 304] (5.3, 2);
\draw[line width = 0.5, blue] (5.47, 1.5) to [out = 110, in = 305] (4.75, 2.55);
\draw[line width = 0.7, red] (5.5, 1.5) to [out = 110, in = 300] (4.5, 2.95);
\draw[->, line width = 0.5, dashed, red] (2, 3) to [out = -20, in = 190] (4.55, 2.7);
\draw[->, line width = 0.5, dashed, blue] (2, 3) to [out = -20, in = 190] (4.8, 2.3);
\draw[->, line width = 0.5, dashed] (2, 3) to [out = -20, in = 190] (5.2, 1.8);
\node at  (2.7, 3.2) {\tiny  \it homotopies $ \g_\l = \{\g^{(U_\l(s))}\}$};
\node[red] at  (5, 3) {$\g^{(U)}$};
\draw[fill, red]  (4.5, 2.95) circle [radius = 0.04];
\draw[fill, blue]  (4.77, 2.55)circle [radius = 0.04];
\draw[fill]  (5.3, 2) circle [radius = 0.04];
\draw[ blue, ->, line width = 0.7]  (6, 2.55) to [out=10, in=130] (9,1);
\draw[->, blue, line width = 1]  (8,0.7) to (11,0.7);
\draw[fill, blue]  (8, 0.7) circle [radius = 0.05];
\node[blue] at  (8, 0.5) {\tiny $0$};
\node[blue] at  (11, 0.5) {\tiny $+\infty$};
\node[blue] at  (9.3, 2.3) {\small $\bW(\l) \= \cW^{(U_o, U_\l, \g_\l)}$};
\node[magenta] at (3,1.4) {\small the space of curves $\cG$};
\node[magenta] at (6.85,0.7) {\small $\p \cG $ };
 \end{tikzpicture}
 \hskip 3 cm 
 }
   \centerline{\tiny\bf Fig. 7}
\medskip
We call this function the   {\it labour function} of the homotopy $\g(t, s) = \g^{(U(s))}(t)$. Being $\bW(0) = 0$,  we can  formulate the Principle of Minimal Labour in the following equivalent form. \\[10pt]
{\bf Principle of Non-Negative  Labour Functionals.} {\it Necessary and sufficient  condition for an element  $U_o \in \wt \cK'$ to be a a $p$-optimal solution  is that for each pair $(U, \g)$ as above, the corresponding  labour function $\bW(\l)$
has a minimum at $\l = 0$.}\par
\bigskip
As immediate consequence of  this statement is that {\it $U_o = (u_o(t), (a_o, b))$ is a $p$-optimal solution  (thus an optimal one) only if
\beq \label{vertical}  \frac{d\bW}{d \l}\bigg|_{\l = 0}  =  - \iint_{t\in [0,T], s\in [0,1]}{\cH}_{u^\ell }\big|_{(\g^{(U_o)}(t), u_o(t))} Y^\ell \big|_{(\g^{(U_o)}(t), u_o(t))} dt\, ds \geq 0\ .\eeq
for any of the vector fields   of the form $Y = Y^\ell\frac{\p}{\p u^\ell}$, which are  defined  at the points $(\g^{(U_o)}(t), u_o(t))$ as  
\beq \label{Ydef} Y \= \frac{\p u^{\ell}(t, s')}{\p s'} \bigg|_{(t, s' =0)}\frac{\p}{\p u^\ell}\big|_{(\g^{(U_o)}(t), u_o(t))}\eeq
for some homotopy $\g(t, s) = \g^{(U(s))}(t)$ corresponding to a   curve $U(s) \in \wt \cK'$ originating from $U_o$.}\par
\medskip
In the  special situations, in which the  homotopies $\g(t, s) = \g^{(U(s))}(t)$ of the above kind  are so many that, by means of \eqref{Ydef},  they generate all possible vector fields of the form $Y =  Y^\ell\frac{\p}{\p u^\ell}$ at the points of the curve $(\g^{(U_o)}(t), u_o(t))$ (and thus also the opposite   $- Y = - Y^\ell\frac{\p}{\p u^\ell}$),  condition \eqref{vertical}
holds  if and only if  condition
\beq \label{vertical1}  \frac{\p \cH}{\p u^\ell}\bigg|_{(t, q(t), p(t), u_o(t))} = 0 \qquad \text{for each} \ t \in [0,1]\eeq
is satisfied.\par
\smallskip
We remark that  condition \eqref{vertical} (and its consequent pointwise  version \eqref{vertical1}) can be also  obtained as a direct consequence  of the classical Pontryagin Maximum Principle. In fact, \eqref{vertical} is sometimes  considered as an infinitesimal version of such principle. However, the above line of   arguments, which are independent  of the classical PMP,  makes clear the fact  that, on the contrary,  \eqref{vertical} and the classical PMP 
are indeed  two independent   consequences of the same  necessary and sufficient criterion,  namely of  the Principle of Minimal Labour. \par
\smallskip
Note also that  \eqref{vertical}  and the PMP are   obtained by taking the  first derivatives of   two different functionals: the first comes from derivatives of  the 
Labour Functionals $\bW(\l)$,  determined by  the  homotopies $\g(t, s)$;  the second comes from the functionals $W(\ve)$,  determined 
by  one-parameter families of needle variations $u_\ve = u^{(\t, \o, \ve)}$ and, consequently,  by  the {\it one-parameter families}  of  homotopies $\g_\ve = \{\g_\ve(\cdot,s), s \in [0,1]\}$ (distinguished one from the other by the independent variable $\ve$)  defined in   \eqref{uks} .
\par
\medskip
\subsection{A few  lines  for   further developments}\hfil\par
\label{suggestions}
\medskip
Suppose that a pair $U_o = (u_o, (a_o, b)) \in \wt \cK'$ is such that 
\beq \label{vertical2}  \frac{\p \cH}{\p u^a}\bigg|_{(t, q(t), p(t), u_o(t))} = 0 \qquad \text{for each} \ t \in [0,1]\ .\eeq
Hence it trivially satisfies the necessary condition \eqref{vertical1}.  The same ideas that led to   \eqref{vertical1} now    imply that $U_o$ is optimal only if,  for any homotopy $\g(t,s)$ as above, the {\it second derivative at $\l = 0$  of the labor functional $\bW$ is non-positive}. \par
\smallskip
Following this pattern,  a whole  sequence of   necessary conditions  could be obtained just   by   looking at  the  derivatives of  higher order of the functional $\bW$,  namely of  order three, four and so on.   
\par
\medskip
It is noteworthy to point out that several important generalized high order conditions have been already  determined in the literature (see e.g. \cite{LS, Su}).  From what we can see at the moment,  such conditions seem to be  strictly   related with the necessary conditions that one can obtain from  labour functionals   by considering higher order derivatives at $\l = 0$. An investigation of such higher order derivatives should necessarily include a   careful comparison  with  the  higher order conditions existing in the literature. \par
\par
\bigskip
Another area of studies is suggested by the analogies and differences between 
the proofs  of the classical PMP and the first order condition \eqref{vertical} or \eqref{vertical1}. Roughly speaking, 
they are both obtained by taking  first derivatives of two   functionals,  of very different construction one from the other. Thus  it might be worth  to make a comparative study also  of the    (generalised) higher order derivatives of such two involved functionals.
\par
\bigskip
Before concluding, we would like to point out that  an analogue of the Principle of Minimal Labour exists for any control problem, 
in which the  correspondence $U \in \cK \longmapsto \g^{(U)} \in \cG$ 
is given by associating to an appropriate control parameter $U$  the solution $\g^{(U)}$ of 
Euler-Lagrange equations, determined by time-dependent Lagrangians of  higher order  and  with  cost functional 
depending on high order derivatives (\cite{CS1}). 
\par

\font\smallsmc = cmcsc8
\font\smalltt = cmtt8
\font\smallit = cmti8
\hbox{\parindent=0pt\parskip=0pt
\vbox{\baselineskip 9.5 pt \hsize=3.1truein
\obeylines
{\smallsmc 
Franco Cardin
Dipartimento Matematica Tullio Levi-Civita
Universit\`a degli Studi di Padova
Via Trieste 63
I-35121 Padova
ITALY
}\medskip
{\smallit E-mail}\/: {\smalltt cardin@math.unipd.it}
}
\hskip 0.0truecm
\vbox{\baselineskip 9.5 pt \hsize=3.7truein
\obeylines
{\smallsmc
Andrea Spiro
Scuola di Scienze e Tecnologie
Universit\`a degli Studi di Camerino
Via Madonna delle Carceri 9A
I-62032 Camerino (Macerata)
ITALY
}\medskip
{\smallit E-mail}\/: {\smalltt andrea.spiro@unicam.it}
}
}

\end{document}